\documentclass[3p,times,procedia]{elsarticle}
\usepackage{nupha_ecrc}
\usepackage{hyperref}

\volume{00}

\firstpage{1}

\journalname{Nuclear Physics A}

\runauth{}

\jid{nupha}

\jnltitlelogo{Nuclear Physics A}

\usepackage{amssymb}

\usepackage[figuresright]{rotating}

\begin{document}

\begin{frontmatter}



\dochead{XXVIIth International Conference on Ultrarelativistic Nucleus-Nucleus Collisions\\ (Quark Matter 2018)}

\title{Non-strange and strange D-meson and charm-baryon production in heavy-ion collisions measured with ALICE at the LHC}


\author{Xinye Peng, for the ALICE collaboration}

\address{Central China Normal University, China, 

Universit$\grave{a}$ degli Studi di Padova $\&$ INFN, Padova, Italy}

\begin{abstract}
We present recent results on strange and non-strange D-meson production measured with ALICE in Pb--Pb collisions at the LHC. In addition, the measurements of the  $\Lambda_{{\rm c}}^+$-baryon production and of the $\Lambda_{{\rm c}}^+/{\rm D^0}$ ratio in pp, p--Pb, and, for the first time ever, Pb--Pb collisions are reported.
\end{abstract}


\end{frontmatter}


\section{Introduction}
Heavy quarks are produced in hard-scattering processes over short time scales compared to the Quark Gluon Plasma (QGP) formation time. They probe the whole system evolution via interact with the medium constituents. The investigation of charm production in Pb-Pb collisions helps understand the colour-charge and quark-mass dependence of in-medium energy loss, the sensitivity of charm quarks to the medium collective motion, whether they reach thermal equilibrium, and charm-hadron production mechanism. For the latter a possible enhancement of ${\rm D_s}/{\rm D^0}$ and $\Lambda_{{\rm c}}/{\rm D^0}$ ratios is predicted by models including charm-hadron formation via coalescence~\cite{cls1}. In pp and p--Pb collisions, the aforementioned particle ratios allow investigating further the charm hadronisation mechanism at LHC energies. In p--Pb collisions the study of the $\Lambda_{{\rm c}}/{\rm D^0}$ ratio allows investigating whether nuclear matter effects on charm hadronisation could enhance the baryon-to-meson ratio with respect to pp collisions also in small systems, as observed in the light flavour sector.
\section{Heavy-flavour reconstruction strategy}
The analysis data sample consists of $370\cdot10^6$ pp at $\sqrt{s}$ = 7 TeV (corresponding to $L_{\rm int}$ = 6.0/nb), $600\cdot10^6$ p--Pb (corresponding to $L_{\rm int}$ = 292/nb) and $100\cdot10^6$ Pb--Pb at $\sqrt{s_{\scriptscriptstyle \rm NN}}$ =5.02 TeV (corresponding to $L_{\rm int}$ = 13.4/nb) minimum-bias collisions, collected by ALICE~\cite{alice} in 2010, 2016 and 2015, respectively. Charmed hadrons are reconstructed at mid-rapidity via the hadronic decay channels ${\rm D}^{0} \to {\rm K}^-\pi^+$, ${\rm D}^+ \to {\rm K}^-\pi^+\pi^+$, ${\rm D}^{*+} \to {\rm D}^0 \pi^+$, ${\rm D}^+_{\rm s} \to \phi\pi^+$, $\Lambda_{{\rm c}}^+ \to {\rm p}{\rm K}^-\pi^+$ and $\Lambda_{{\rm c}}^+ \to {\rm p}{\rm K}^0_{\rm s}$. To reduce the combinatorial background, selections on the decay topology and particle identification are applied. In the case of $\Lambda_{{\rm c}}^+$ analysis, a second approach based on multivariate approach (BDTs) to select decay topology is used. The signal is extracted via an invariant-mass analysis. The feed-down from beauty-hadron decays is subtracted according to expectations based on FONLL calculations, and in p-Pb and Pb-Pb collisions, with further assumptions of feed-down nuclear modification factor ($R_{\rm AA}^{\rm feed-down}/R_{\rm AA}^{\rm prompt}$ = 1 in p--Pb, and ranges from 1 to 2 in Pb--Pb collisions depending on $p_{\rm T}$, centrality classes and particle species). 
 \section{Results}
   \begin{figure}[!htbp]
  \begin{center}
  \includegraphics[width=.33\textwidth]{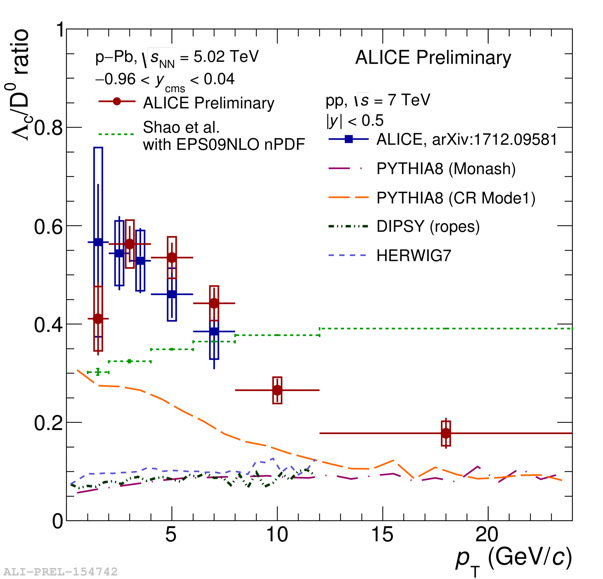}
  \includegraphics[width=.66\textwidth]{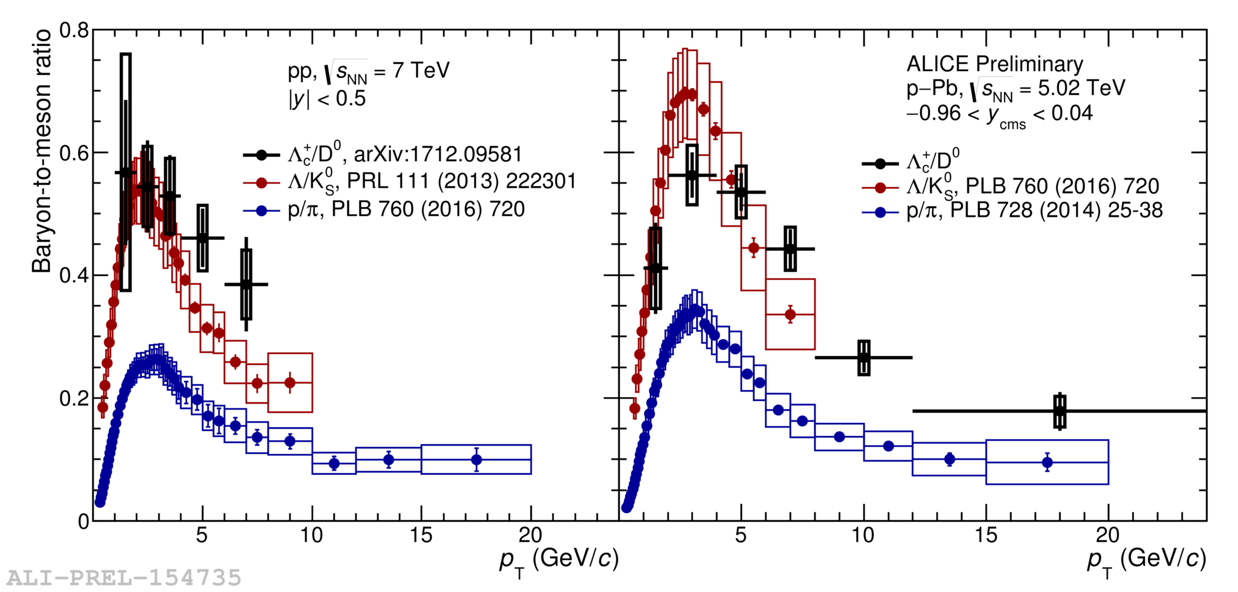}
  \end{center}
  \caption{The $\Lambda_{{\rm c}}/{\rm D^0}$ raito in pp collisions at $\sqrt{s}$ = 7 TeV and in p--Pb collisions at $\sqrt{s_{\scriptscriptstyle \rm NN}}$ = 5.02 TeV, compared with model predictions (left) and $\Lambda/{\rm K}^0_{\rm s}$, ${\rm p}/{\pi}$ ratios in pp (middle) and p--Pb (right) collisions.  }
  \label{fig:lambdacppb}
  \end{figure}
 In this contribution, new results on the $p_{\rm T}$-differential cross section of the $\Lambda_{{\rm c}}^+$-baryon, measured in p--Pb collisions at $\sqrt{s_{\scriptscriptstyle \rm NN}}$ = 5.02 TeV with the data collected in 2016 are presented. The precision is improved and $p_{\rm T}$ coverage is extended with respect to the previous measurement~\cite{lambdac1}. Figure~\ref{fig:lambdacppb} shows the updated baryon-to-meson ratio $\Lambda_{{\rm c}}^+/{\rm D^0}$. The ratios obtained in pp and p--Pb collisions are compatible within the larger uncertainties of the pp measurements. On the left panel of Figure~\ref{fig:lambdacppb}, the measured $\Lambda_{{\rm c}}^+/{\rm D^0}$ ratios are compared with models. The results are higher than the expectation from theoretical models including PYTHIA8 with Monash tune and with a tune with enhanced colour reconnection~\cite{pythia8}, DIPSY with ropes~\cite{dipsy}, HERWIG7 with a cluster hadronisation mechanism~\cite{herwig7}, and a calculation~\cite{Lansberg:2016deg} tuned on LHCb pp data~\cite{Aaij:2013mga} at forward rapidity, note that this calculations is higher than data at high $p_{\rm T}$. PYTHIA8 with enhanced colour reconnection gets closer to the data, hinting the importance of understanding the role of colour reconnection in charm hadronisation. On the right panel, the baryon-to-meson ratio in the charm sector ($\Lambda_{{\rm c}}^+/{\rm D^0}$) is compared with the same ratios in the light flavour sectors~\cite{ratio1,ratio2,ratio3}($\Lambda/{\rm K}^0_{\rm s}$ and ${\rm p}/{\pi}$). A similar trend is observed with decreasing values from $p_{\rm T}$ = 4 $\mathrm{GeV}/c$.
   \begin{figure}[!htbp]
  \begin{center}
  \includegraphics[width=.49\textwidth]{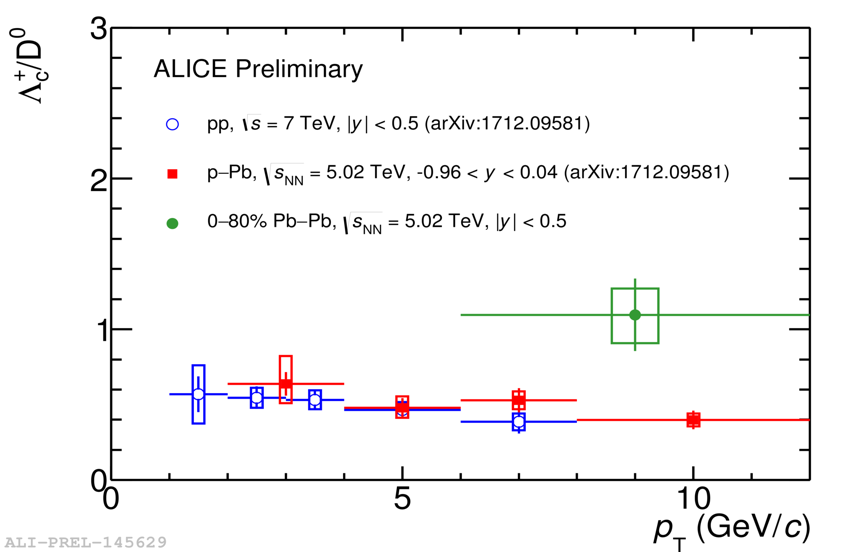}
  \includegraphics[width=.49\textwidth]{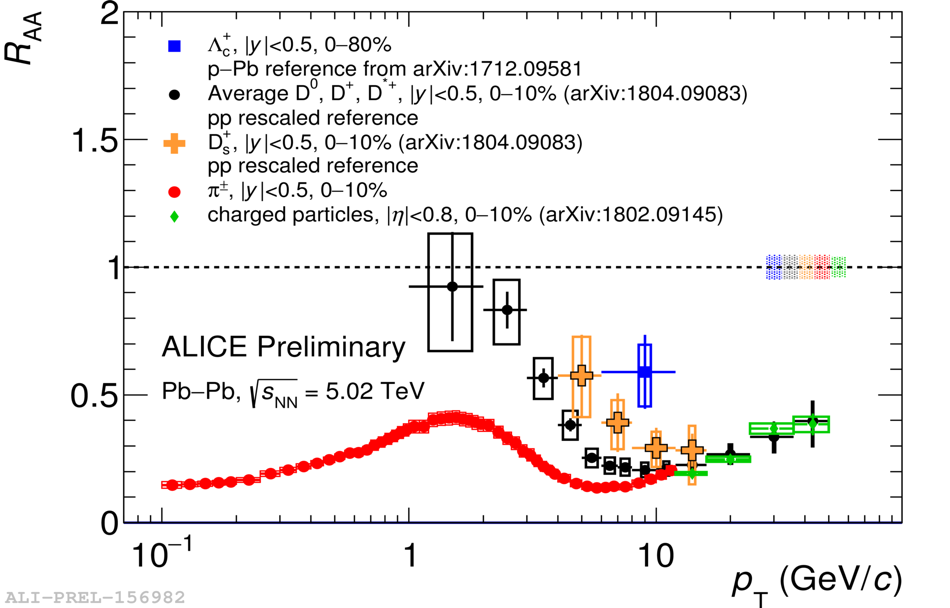}
  \end{center}
  \caption{The $\Lambda_{{\rm c}}^+/{\rm D^0}$ ratio in Pb--Pb collisions at $\sqrt{s_{\scriptscriptstyle \rm NN}}$ = 5.02 TeV, compared with that in pp and p--Pb collisions (left), and $\Lambda_{{\rm c}}^+$-baryon $R_{\rm AA}$ in 0-80\% Pb--Pb collisions at 5.02 TeV compared to average non-strange D-meson, ${\rm D}_{\rm s}^+$-meson, $\pi ^\pm$ and charged particle $R_{\rm AA}$ in 0-10\% centrality class}
  \label{fig:lambdacpbpb}
  \end{figure}
  
 At this conference, ALICE presented the first measurement of the $\Lambda_{{\rm c}}^+/{\rm D^0}$ ratio in Pb--Pb collisions. The measurement is performed at $\sqrt{s_{\rm NN}}$ = 5.02 TeV in the centrality class 0-80\% for 6 $<$ $p_{\rm T}$ $<$ 12 $\mathrm{GeV}/c$. On the left panel of Figure~\ref{fig:lambdacpbpb}, the comparison of the $\Lambda_{{\rm c}}^+/{\rm D^0}$ ratios in the three colliding systems is reported. It shows a hint of enhancement in Pb--Pb with respect to pp and p--Pb collisions. The value is similar to that measured by STAR at lower $p_{\rm T}$ in Au--Au collisions at 200 GeV~\cite{star}. The models~\cite{cls1,model1,model2,model3} tend to underestimate the data for the same $p_{\rm T}$ interval. The right panel of Figure~\ref{fig:lambdacpbpb} shows the $\Lambda_{{\rm c}}^+$-baryon $R_{\rm AA}$ measured in 0-80\% Pb--Pb collisions at 5.02 TeV compared to the average non-strange D-meson~\cite{D0}, ${\rm D}_{\rm s}^+$-meson~\cite{D0}, $\pi ^\pm$  and charged particle~\cite{charge} $R_{\rm AA}$ in 0-10\% centrality class. A hint of hierarchy is observed: $\Lambda_{{\rm c}}^+$-baryon $R_{\rm AA}$ $>$ ${\rm D}_{\rm s}^+$-meson $R_{\rm AA}$ $>$ non-strange D-meson $R_{\rm AA}$, which reflects the enhancement of baryon and strange-particle production expected by models including charm hadronisation via coalescence of charm quarks with the surrounding quarks in the QGP. For $p_{\rm T}$ $>$ 10 $\mathrm{GeV}/c$, non-strange D-meson $R_{\rm AA}$ is similar to $\pi ^\pm$ $R_{\rm AA}$, in agreement with what expected from the combination of the colour charge and mass dependence of energy loss with the different fragmentation and initial spectra of charm and light partons~\cite{energy}. For $p_{\rm T}$ $<$ 8 $\mathrm{GeV}/c$, non-strange D-meson $R_{\rm AA}$ is higher than that of $\pi ^\pm $. This difference does not prove per se a smaller energy loss for charm quarks than light quarks: other effects must be considered, like $N_{{\rm part}}$ scaling of pion production at very low $p_{\rm T}$, the different impact of radial flow, coalescence, as well as different initial-state effects.
  \begin{figure}[!htbp]
  \begin{center}
  \includegraphics[width=.49\textwidth]{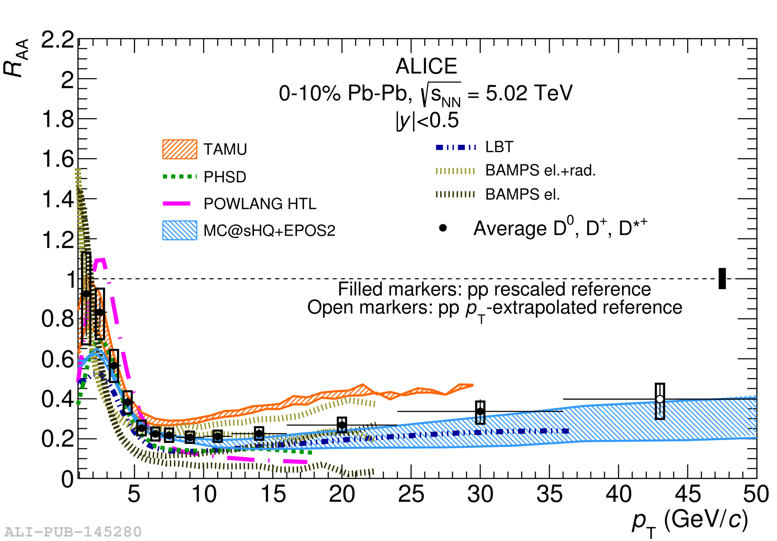}
  \includegraphics[width=.49\textwidth]{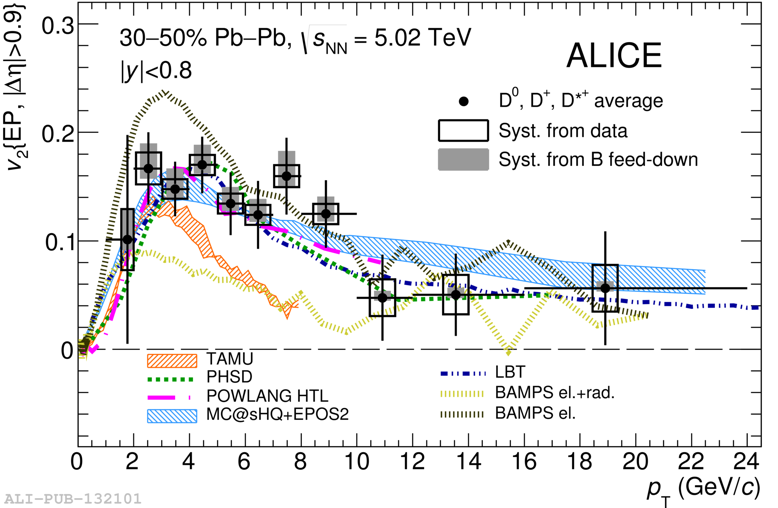}
  \end{center}
  \caption{Non-strange D-meson $R_{\rm AA}$ in the 0-10\% centrality class~\cite{D0} (left) and elliptic flow $v_{2}$ in the 30-50\% centrality class~\cite{v2} (right), compared with models predictions~\cite{LBT,MCsHQ,PHSD,POWLANG,Uphoff:2014hza,He:2014cla}.}
  \label{fig:raav2}
  \end{figure}
 
The simultaneous comparison of both $R_{\rm AA}$ and elliptic flow ($v_{2}$) can provide important constraints on the theoretical models and help to extract information about the medium properties. Figure~\ref{fig:raav2} shows the non-strange D-meson  $R_{\rm AA}$ in the 0-10\% centrality class (left) and elliptic flow $v_{2}$ in the 30-50\% centrality class (right)~\cite{v2}, compared with the  models. Models (LBT~\cite{LBT}, MC@sHQ~\cite{MCsHQ}, PHSD~\cite{PHSD}, POWLANG~\cite{POWLANG}) in which charm quarks pick up collective flow via recombination or subsequent elastic collision in the expanding QGP medium can better describe both $R_{\rm AA}$ and $v_{2}$ at low $p_{\rm T}$. The models which can reasonably describe the data use a diffusion coefficient $2\pi{T}{D}_{s}(T)$ in the range of 1.5-7, corresponding to a charm thermalisation time $\tau_{\rm charm}$ in the range of 3-14 $\mathrm{fm/c}$ at $T_{c}$.

\section{Summary}
The ALICE results on charm-meson and charm-baryon production have been reported. The $\Lambda_{{\rm c}}^+/{\rm D^0}$ baryon-to-meson ratio in pp and p--Pb collisions is higher than theoretical predictions, and a similar $p_{\rm T}$-trend is observed compared to $\Lambda/{\rm K}^0_{\rm s}$ and ${\rm p}/{\pi}$ ratios. The first LHC measurement of $\Lambda_{{\rm c}}^+/{\rm D^0}$ ratio in Pb--Pb collision at 5.02 TeV shows a hint of enhancement with respect to pp and p-Pb collisions. Finally, the ${\rm D}^+_{\rm s}$-meson $R_{\rm AA}$ is higher than non-strange D-meson $R_{\rm AA}$ supporting the possibility of charm hadronisation via recombination. A significant positive D-meson $v_{2}$ confirms that charm quark is sensitive to the medium collective motion.

\section{Acknowledgement}
This work was partly supported by the National Key Research and Development
Program of China under Grant No. 2016YFE0100900 and the NSFC
(Grant Nos. 11775095 and 11475068) and Grant No. CCNU18ZDPY04.




\bibliographystyle{utphys}
\bibliography{xinye}







\end{document}